\documentclass[12pt,preprint]{aastex}
\usepackage{color}

\shorttitle{EUV Wave and Prominence Activation}
\shortauthors{Takahashi et al.}

\begin{document}
\title{Prominence Activation by Coronal Fast Mode Shock}

\author{
Takuya Takahashi\altaffilmark{1,3},
Ayumi Asai\altaffilmark{2},
Kazunari Shibata\altaffilmark{3}}

\email{takahashi@kwasan.kyoto-u.ac.jp}
\altaffiltext{1}{
Department of Astronomy, Kyoto University,
Sakyo, Kyoto, 606-8502, Japan}
\altaffiltext{2}{
Unit of Synergetic Studies for Space, Kyoto University,
Yamashina, Kyoto 607-8471, Japan.}
\altaffiltext{3}{
Kwasan and Hida Observatories, Kyoto University,
Yamashina, Kyoto 607-8471, Japan.}

\begin{abstract}
An X5.4 class flare occurred in active region (AR) NOAA11429 on 2012 March 7. The flare was associated with very fast coronal mass ejection (CME) with its velocity of over 2500~km~s$^{-1}$. 
In the images taken with {\it STEREO-B}/COR1, a dome-like disturbance was seen to detach from expanding CME bubble and propagated further.
A Type-II radio burst was also observed at the same time. 
On the other hand, in EUV images obtained by {\it SDO}/AIA, expanding dome-like structure and its foot print propagating to the north were observed.
The foot print propagated with its average speed of about 670~km~s~$^{-1}$ and hit a prominence located at the north pole and activated it. While the activation, the prominence was strongly brightened. On the basis of some observational evidence, we concluded that the foot print in AIA images and the ones in COR1 images are the same, that is MHD fast mode shock front. 
With the help of a linear theory, the fast mode mach number of the coronal shock is estimated  to be between 1.11 and 1.29 using the initial velocity of the activated prominence. Also, the plasma compression ratio of the shock is enhanced to be between 1.18 and 2.11 in the prominence material, which we consider to be the reason of the strong brightening of the activated prominence. The applicability of linear theory to the shock problem is tested with nonlinear MHD simulation.

\end{abstract}


\section{Introduction}
The solar corona is a place where plasma transient phenomena such as flares, coronal mass ejections (CMEs) and jets are continuously observed. These are vigorously investigated in the framework of magnetohydrodynamics (e.g. \citet{shibm2011}). When magnetic energy is locally liberated in a short time scale as in flares or jets, magnetohydrodynamic shock waves are generated. The shock waves sometimes propagate into the interplanetary space and reach the Earth. The arrival of the shock waves to the Earth sometimes causes the sudden increase of radiation level and sudden change of Earth's magnetic field by compressing it.

The existence of coronal disturbances associated with flares are suggested by some observation of far-away filament oscillation after flares. This is called 'winking filament' after the observational feature where the filament appeared and disappeared in H$\alpha$ \citep{smit1971}. Sometimes, Type II solar radio bursts are observed associated with solar flares. Type II radio bursts are observed in dynamic radio spectra as bands slowly drifting down often in pairs differing in frequency by a factor of about 2, and interpreted as manifestation of coronal shock propagation \citep{wild1954,kai1970}.

Globally propagating disturbances associated with solar flares are first observed in H$\alpha$. A fan-shaped 'wave fronts' were observed to propagate in the chromosphere after the occurrence of a flare \citep{more1960}. This phenomenon was named 'Moreton wave' after the name of the discoverer. The wave front of Moreton waves appear bright in line center and blue wing and dark in red wing followed by a fainter front whose signature of the intensity change is reversed (dark in the blue wing and bright in the red wing). This observational feature suggest down-up swing of chromospheric plasma during the passage of Moreton wave. Moreton wave propagated with speeds between 500~km~s~$^{-1}$ and 1500~km~s~$^{-1}$, and lasted for about ten minutes. In some events, Moreton wave front propagated faster than 2000~km~s~$^{-1}$. The sound velocity in the chromosphere is about 10~km~s$^{-1}$. Fast mode magnetosonic wave speed is also the same order. If Moreton waves are magnetohydrodynamic wave propagating in the chromosphere, they have to be a strong shock wave with the Mach number of over 10. It was difficult to accept that such strong shock waves propagate a long distance without any significant dissipation.

\citet{uchi1968} gave a comprehensive explanation of Moreton waves as a cross section between coronal MHD fast mode shock front and the chromosphere. This model of Moreton waves was widely accepted as 'Uchida's Moreton wave model' soon. According to Uchida model, there has to be a coronal counterpart of Moreton wave (namely coronal MHD fast mode shock front). People call the expected coronal counterpart of Moreton wave as 'invisible' Moreton wave. Winking filament were discussed as a result of the passage of invisible Moreton waves \citep{smit1971,eto2002,okam2004}.

In 1990's, wave-like disturbances propagating in the corona after flares were observed with the Extreme Ultra Violet (EUV) Imaging Telescope (EIT; Delabourdiniere et al. 1995) on board {\it Solar and Heliospheric Observatory} ({\it SOHO}; Domingo et al. 1995). This wave-like phenomenon was called 'EIT wave' after the name of the observational instrument. The EIT wave front propagated concentrically from the flare site with the speed of about 250~km~s~$^{-1}$ \citep{thom1999,thom2000}.
At first, EIT waves were thought to be the expected coronal counterparts of Moreton waves. 
However, EIT waves and Moreton waves showed quite different observational features \citep{eto2002,shib2011,zhan2011}, which leads to a long standing discussion of the physical nature of EIT waves\citep{warm2001,vrsn2002,trip2007,warm2007, will2009, gall2010}.

Recently, with the Atmospheric Imaging Assembly (AIA; \citet{tit2006, lem2011}) on board the {\it Solar Dynamic Observatory} ({\it SDO}; Pesnell et al. 2012), fast coronal waves have been observed associated with flares \citep[e.g.][]{liu2010,ma2011}. \citet{chen2011} found two different coexisting coronal waves, a slow coronal wave (i.e. EIT wave) and a fast coronal wave. These coronal disturbances are generally called EUV waves. \citet{asai2012} reported a first simultaneous observation of H$\alpha$ Moreton wave, fast EUV wave and filament/prominence oscillations. In that paper, they showed that the fast EUV wave was the true coronal counterpart of the H$\alpha$ Moreton wave. They also reported that after the Moreton wave disappeared, fast EUV wave propagated further and activated large amplitude filament/prominence oscillation.

Large amplitude prominence oscillations caused by EUV waves are often observed by {\it SDO}/AIA. Such prominence oscillations are generally driven in a very short time scale and behave as damping oscillators, while in some cases, the oscillations lead to the eruption of the prominences. They are sometimes driven by a sudden change of magnetic field structure through magnetic reconnections \citep{vrsn2007}. These oscillations have been investigated in the context of diagnosis of physical quantities and eruptive instability of the prominences \citep{isob2007,trip2009}. 
The prominence activations by coronal shock waves and its subsequent oscillation provide us with the insight into physical properties of both coronal shock waves and the prominences (such as the width of the wave front and magnetic field structure supporting the prominences) \citep{gilb2008,liu2013}.

In this study, we analyzed the prominence activation process during the collision with an EUV wave associated with an X5.4 class flare on 2012 March 7. 
The unprecedented time and spatial resolution of multi-wavelength EUV data taken with SDO/AIA enables us to follow the prominence activation process in detail.
We concluded, based on some observational evidence, that the EUV wave that activated the prominence is the same as the coronal shock wave propagating ahead of CME flank, which was seen in the images taken with the inner Coronagraph (COR1) of the Sun Earth Connection Corona and Heliospheric Investigation (SECCHI) on board the {\it Solar Terrestrial Relations Observatory} ({\it STEREO}; Kaiser et al. 2008; Dreasman et al. 2008)-{\it Behind} satellite. We could explain the compression and acceleration of activated prominence as a result of coronal shock wave transmission into the dense prominence. Furthermore, we estimated the coronal shock strength using initial velocity of activated prominence.

In section 2, we overview the observation of X5.4 class flare and associated coronal disturbances. In section 3, we estimate the coronal shock strength using the initial velocity of activated prominence.

\section{Observation and analysis}
\subsection{Overview of the flare and CME}
Figure 1 shows the soft X-ray light curve taken with {\it GOES} satellite. The first X5.4 class flare occurred in active region (AR) NOAA11429 at 00:02UT and peaked at 00:24UT on 2012 March 7. One hour after the onset of the first flare, the second X1.3 class flare occurred at different part of the same AR. The two flares each were accompanied by fast coronal mass ejections (CMEs). The velocity of each CME is reported to be 2684~km~s~$^{-1}$ and 1825~km~s~$^{-1}$, respectively.\footnote{see,
http://cdaw.gsfc.nasa.gov/CME\_list/UNIVERSAL/2012\_03/univ2012\_03.html}
The abrupt change of horizontal magnetic field and perpendicular Lorentz force near the polarity inversion line (PIL) during X5.4 class flare are reported, and relationship between the abrupt change and the fast propagation velocity of the CME is pointed out \citep{pet2012,wang2012}. In this study we investigate the coronal disturbances associated with the first flare of the two. 

\subsection{A disturbance ahead of CME flank and Type II radio burst}
On March 7, the AR NOAA11429 was located at (N17,E16) in the heliographic coordinate. The {\it STEREO-B} satellite was then observing the Sun at 118 degrees east of the Earth having captured the CME just from the side. We analyzed the 5 minutes cadence data taken with {\it STEREO-B}/SECCHI/COR1.

Figure 2 shows the images taken with COR1 and EUVI on board the {\it STEREO-B} satellite. Figure 2(a) is the image taken with extreme ultraviolet imager (EUVI; W\"{u}lser et al. 2004)  304~{\AA}  pass band. In the EUVI 304~{\AA}  image, we can see the polar prominence clearly. Figure 2(b) to (d) are the difference image of COR1 image with embedded image of Figure 2(a).  In Figure 2(b), a disturbance propagating ahead of expanding CME plasma appeared at 00:25UT. This disturbance reached the sky just over the polar prominence at 00:31UT, and propagated further as shown in (c) and (d). The average speed of propagation of the disturbance from 00:25UT to 00:36UT was 420~km~s~$^{-1}$ measured along the limb of the occulting disk of the COR1 images.

On the other hand, in the dynamic spectrum in the range between 25MHz and 2500MHz taken with the Hiraiso Radio Spectrograph (HiRAS, Kondo et al. 1995), Type II radio burst was observed. The second harmonic component of the Type II radio burst appeared at 00:19:12UT in 200MHz and disappeared at 00:30:36UT in 50MHz. From the average drift rate of the Type II radio burst, we can estimate the radial component of propagation velocity of the shock front ahead of the CME to be 672~km~s~$^{-1}$ \citep{new1961,mann1999}.  Because of the coincidence, we regarded the coronal disturbance ahead of the CME bubble in the COR1 images to be the foot print of the shock wave which caused the Type II radio burst.

\subsection{EUV observation}
In order to investigate the time evolution of the coronal disturbances, we used the extreme ultra violet (EUV) images taken with {\it SDO}/AIA. In this study, we used 2 pass band of AIA, namely 193~{\AA}  and 304~{\AA} . The time cadence of AIA data is 12 seconds while we used 24 seconds cadence data in our analysis.

Figure 3 shows the flare associated disturbances in 193~{\AA}  images. At 00:18UT, dome-like structure appeared over the flaring AR. This dome-like structure kept on expanding and propagated upward from the solar disk. On the other hand, strong disturbance appeared on the north east of the AR at 00:23UT and propagated in the lower corona toward the north pole, and hit a polar prominence located there. By comparing COR1 and AIA data, we concluded that dome-like structure seen in AIA 193~{\AA} images is identical to the shock front ahead of CME seen in COR1 images. We consider the foot print propagated in the lower corona in AIA 193~{\AA} images are the foot print of the shock (Figure 4).

These disturbances seen in EUV are both called 'EUV waves'. In the following, we distinguish the two disturbances as 'dome-like structure' and 'foot print' respectively. The dome-like structure expanded further and left away from the AIA field of view at 00:27UT. The foot print propagated further and hit the polar prominence at 00:33UT and caused large amplitude prominence oscillation. The average propagation velocity of the foot print between 00:18UT and 00:33UT was 670km~s~$^{-1}$. We can also see that the wave front of the foot print more and more inclined to the propagation direction during its propagation. 

In Figure 3(b), the dome-like structure seems to already reach the polar prominence at 00:23UT. However, we notice that this is just because of a projection effect of three dimensionally expanding dome by comparing it with the shock front seen in STEREO-B/COR1 images.

\subsection{Prominence activation}
Taking a close look at the AIA 304~{\AA} images, we can clearly see how the large amplitude prominence oscillation was activated. Before the foot print 'hit' the prominence at 00:33UT, the prominence was slowly swaying (Figure 5(a)). At 00:33UT, the foot print of the coronal shock wave hit the prominence (Figure 5(b)). When it hit the edge of the prominence, the prominence was strongly brightened. The brightness was about three times of that part before hit by the disturbance. After that the bright part propagated to the right in the image. The prominence was brightened from 00:33UT to 00:37UT, and during this time, the prominence material was accelerated from east to west in series (Figure 5(c)). 

We can also see that the prominence was accelerated perpendicular to the wave front in AIA 193~{\AA} images. In 193~{\AA} images, the prominence was seen as dark prominence (Figure 6(a)-(d)). It was brightened just when the foot print arrived but not a long or strong brightening could be seen as in 304~{\AA} images. From AIA 193~{\AA} images, the propagation velocity of the wave front of the foot print just before the injection to the prominence was 380km~s~$^{-1}$ and the initial velocity of the activated prominence was 48~km~s$^{-1}$.

\section{Estimation of shock strength in the corona}
We concluded that the foot print that activated the polar prominence was the shock wave which propagated ahead of the CME seen in {\it STEREO-B}/COR1 images. Following are the reasons.

First, the propagation location of foot print agreed well with the location of the shock wave ahead of CME bubble.
The shock wave ahead of the CME appeared in {\it STEREO-B}/COR1 image at 00:26UT while the foot print in AIA 193~{\AA} images appeared at 00:23UT. The timing of its appearance agreed well within the time cadence of {\it STEREO-B}/COR1. The shock wave in COR1 images passed the sky just over the polar prominence at 00:31UT while the foot print in AIA 193~{\AA} images hit the prominence at 00:33UT. We think that the 2 minutes delay is due to the inclination of the wave front towards the propagation direction in the corona.

Second, the wave front of the foot print is inclined towards the propagation direction and the inclination increases with time.We thought this is because of wave refraction towards the photosphere because the phase velocity of MHD fast mode wave would increase with height in the lower corona \citep{gopa2001}. The time evolution of the wave front agrees very well with the prediction of Uchida's model of coronal fast mode shock wave \citep{uchi1968}. 
Because of the gravitational stratification, the plasma density of the lower corona is higher, which made the foot print of the dome-like coronal shock front bright in EUV pass band.

Third, the polar prominence was strongly brightened, and at the same time, accelerated perpendicularly to the wave front.
These behavior can be explained simply as a result of injection of compressive wave into the prominence.

Looking at the prominence activation in the direction perpendicular to the wave front, we can regard this process as one dimensional fast mode shock wave transmission from corona into the dense prominence.
In the linear case, the continuity of energy flux and momentum flux of the wave are written as 
\begin{equation}
\rho_c\left( {V_i}^2-{V_r}^2 \right)C_c=\rho_p {V_t}^2C_p
\end{equation}
\begin{equation}
\rho_c\left(V_i-V_r\right)C_c=\rho_pV_{t}C_p,
\end{equation}
which reduces the equation
\begin{equation}
V_i=\frac{1}{2}\left(1+\frac{\rho_pC_p}{\rho_cC_c}\right)V_t,
\end{equation}
where $\rho$, $V$ and $C$ are plasma density, velocity amplitude of the wave, and phase velocity of MHD fast mode wave, respectively. The subscript i, t and r of $V$ denote injected, transmitted and reflected waves, respectively, and the subscript $c$ and $p$ denote the quantities in the corona and prominence, respectively.

Magnetic field strength around the prominence could be stronger than that of ambient corona, and therefore, the fast mode shock property in the corona could change gradually while approaching the prominence. If so, especially, wave refraction due to the spatial variation of fast mode propagation velocity in the corona around the prominence affects how strong the prominence is activated. In figure 7, we show the time-distance plot of the intensity of AIA 193~{\AA} images along the cutting-line shown in Figure 6 (a) from 00:00UT to 02:00UT. In Figure 7, on the other hand, we can see that the coronal fast mode shock front seen as bright signature approaching the dark filament propagates almost in constant speed. From this, we think that the spatial variation of magnetic field strength around the activated prominence did not affect much the process of prominence activation. Although the density discontinuously changes at the corona-prominence boundary, we think it is still reasonable to assume that the magnetic field strength at the corona-prominence boundary is continuous since the length scale of overall magnetic field structure is much larger than the width of the transition layer between corona and prominence.

From the discussion above, we can express the compression ratio $r_c$ of the coronal shock wave using observed propagation velocity of injected wave $C_c$ and initial velocity of activated prominence $V_t$ as 
\begin{equation}
{r_c}=
\frac{C_c}{C_c-V_i}=
\frac{1}{1-\left(1+\sqrt{a}\right)\frac{V_t}{2C_c}},
\end{equation}
where we express ${\rho_p}/{\rho_c}$ as $a$.
In order to simplify the problem, we treat the transmitted shock wave as a perpendicular one further on.
If the local density gap between the prominence and corona is $\chi$ and the volume filling factor of the prominence is $f_V$, $\rho_p$ can be expressed using $\rho_c$ as $\rho_p=f_V\chi\rho_c+(1-f_V)\rho_c$.

The relation between the fast mode Mach number $M_{f,c}$ and the compression ratio $r_c$ in perpendicular shock case is
\begin{equation}
M_{f,c}=\sqrt{\frac{2r_c\left(\left(2-\gamma\right)r_c+\gamma\left(\beta_c+1\right)\right)}{\left(\left(\gamma+1\right)-r_c\left(\gamma-1\right)\right)\left(\gamma\beta_c+2\right)}}.
\end{equation}
In the linear problem, we also relate the compression ratio $r_p$ in the prominence with the one in the corona $r_c$ as
\begin{equation}
r_p=r_c\frac{2\sqrt{a}}{1+\sqrt{a}}.
\end{equation}
We can see from equation (6) that the compression ratio of the shock wave is enhanced when it is transmitted into a dense prominence.

From AIA 193~{\AA} images, we get $C_c{\sin \theta}=380~{\rm km~s~^{-1}}$ and $V_t{\sin \theta}=48~{\rm km~s~^{-1}}$ where $\theta$ is the angle between wave propagation direction and the line of sight direction. Here we assume the local density gap $\chi$ is $100$, the volume filling factor $f_V$ is between $0.001$ and $0.1$\citep{terr2008,lab2010,mac2010}. This leads to the range of the density gap $\rho_p/\rho_c$ between $1.1$ and $10.9$.
Estimated coronal shock properties are shown in Table 1.

Especially, $M_{f,c}$ and $r_p$ falls into the range between 1.11 and 1.29, and between 1.18 and 2.11 respectively. The brightness of the prominence material in AIA 304~{\AA} images is enhanced by a factor of about 2 during the activation, and it is roughly consistent with the result above if the brightening is due to compression.

\section{1.5 dimensional numerical simulation of fast mode wave transmission}
In order to estimate the coronal shock strength, we applied linear theory of wave transmission. In this section, we show the result of 1.5D MHD simulation of this model.

\subsection{Numerical settings}

We numerically solved 1.5 dimensional ideal MHD equations:
\begin{equation}
\frac{\partial\rho}{\partial t}=-\frac{\partial}{\partial x}(\rho V_x)
\end{equation}

\begin{equation}
\frac{\partial}{\partial t}(\rho V_x)=-\frac{\partial}{\partial x}(\rho V_x^2+p+\frac{1}{8\pi}B_y^2)
\end{equation}

\begin{equation}
\frac{\partial B_y}{\partial t}=-\frac{\partial}{\partial x}(V_xB_y)
\end{equation}

\begin{equation}
\frac{\partial}{\partial t}(\frac{1}{2}\rho V_x^2+\frac{p}{\gamma-1}+\frac{1}{8\pi}B_y^2)=-\frac{\partial}{\partial x}\biggl(V_x(\frac{1}{2}\rho V_x^2+\frac{\gamma}{\gamma-1}p+\frac{1}{4\pi}B_y^2)\biggr)
\end{equation}

The numerical scheme is Harten-Lax-van Leer-Discontinuities (HLLD) approximate Riemann solver \citep{miyo2005} with second-order total variation diminishing (TVD) Monotonic Upstream-Centered Scheme for Conservation Laws (MUSCL) and second order Runge-Kutta time integration with hyperbolic numerical divergence cleaning method\citep{dedn2002}.
The x-axis is uniformly discretized by $N=1000$ grid points. Free boundary is applied as a boundary condition and the numerical box is $x \in [-1.0 , 1.0]$.
The ratio of specific heat $\gamma$ is assumed to be $5/3$.
The unit of length, velocity, time and density in the simulation are $x_c$, $V_{Ac}$, $\tau=x_c/V_{Ac}$ and $\rho_c$, respectively, where $V_{Ac}$ is the Alfven speed. The unit of magnetic field strength is given as $B_c=\sqrt{\rho_cV_{Ac}^2}$.
The values of normalization units are set as a typical ones in the quiet sun corona:
$x_c = 1.0\times10^{10}~cm$, $V_{Ac}= 5.0\times10^2~km~s~^{-1}$, $\tau=x_c/V_{Ac}=2.0\times10^2~s$, $\rho_c=1.0\times10^{-15}~g~cm^{-3}$ and $B_c=\sqrt{\rho_cV_{Ac}^2}=1.6~G$.

Initial conditions are as follows,
\begin{equation}
\rho=\left\{ \begin{array}{ll}
\rho_{c}=1.0~~~(x < 0.5) \\
\rho_{p}~~~~~~~~~~~(0.5 < x) \\
\end{array} \right.
\end{equation}
\begin{equation}
p=p_0
\end{equation}
\begin{equation}
{\bf B}=\left( 0, B_0 \right)
\end{equation}
\begin{equation}
{\bf V}=\left\{ \begin{array}{ll}
\left( 2V_0\sin(\frac{2\pi x}{w}), 0 \right) ~~(|x| < w) \\
\left( 0, 0 \right) ~~(|x| > w) \\
\end{array} \right.
\end{equation}
where $\rho_{c}$ and $\rho_{p}$ denotes the density of the corona and prominence respectively.

Initially, gas pressure and magnetic field are set uniform.
The magnetic field direction is perpendicular to the direction of wave propagation, since we think of perpendicular shock injection problem in this study.
There is a density gap at $x=0.5$, left and right side of which being the corona and the prominence, respectively.
The density gap between the corona and the prominence is resolved by one numerical grid.
The initial wave packet is the superposition of two sinusoidal-shaped wave packets propagating in the opposite directions to each other.
We let the sinusoidal-shaped wave packet propagate in the corona and go into the dense region, 'the prominence'.

For quasilinear cases $V_0=0.01$, and for nonlinear cases $V_0=0.2$.
The width of the wave packet $w$ is set to be $0.05$.
For each case, $\rho_{c}=1$, and $\rho_{p}$ ranges from 2 to 20. $B_c$ is fixed to be $\sqrt{4\pi}$ so that the initial Alfven velocity is unity in all cases.
The initial pressure $p_0$ takes the value $0.025$ or $0.1$, each of which corresponds to plasma beta $\beta=0.05$ and $\beta=0.2$.
The initial distribution of $V_x$ and $\rho$ in quasilinear case ($V_0=0.01$) with $\rho_{p}=20$ and is shown in Figure 8.
While the wave injection into the prominence, some part of the wave energy is transmitted while some part are reflected (Figure 9).
In the nonlinear case, initially sinusoidal-shaped wave packet is deformed by nonlinear sharpening effect, and became shock before the injection into the prominence (Figure 10).

In order to compare quasilinear and nonlinear cases, we define 'velocity transmittance' of the wave ${\rm T_v}$ as follows.
\begin{equation}
{\rm T_v}=\frac{V_{t}}{V_{i}}.
\end{equation}
This is the ratio of the velocity amplitude of transmitted and injected waves (see Figure 9 and 10).
The velocity amplitudes of injected and transmitted waves are measured just before and after the injection, respectively.
Linear analytic expression of this quantity as a function of the density gap ${\rm a=\rho_{p}/\rho_{c}}$ is (from equation (11)),
\begin{equation}
{\rm T_v}=\frac{2}{1+\sqrt{a}}.
\end{equation}
Figure 10 compares the numerical results and analytic expression.
(a) and (b) show the cases with $\beta=0.05$ and $\beta=0.20$, respectively.
From the result shown in Figure 10, we can say that if the strength of the injected shock is not strong, the linear analytic treatment is not so bad.

Then, we define value $R_c$ as follows
\begin{equation}
R_c=\frac{\rho}{\rho_0},
\end{equation}
where $\rho_0$ denotes the initial density distribution.
At the wave front, $R_c$ corresponds to the wave compression ratio.
The distribution of $R_c$ in quasilinear case just before and after the wave injection is shown in Figure 12, and the same plot for nonlinear case is shown in Figure 13.
The time evolution of velocity and compression ratio in quasilinear and nonlinear cases are shown in Figure 14 and 15.
The steep overshoot at $x=0.5$ after the wave injection is the result of the rightward shift of the prominence material, so it does not represent the wave compression ratio.
From the linear theory, the compression ratio of the wave is expected to be enhanced after the transmission into the prominence (equation (6)).
The expected enhancement of the compression ratio after the wave transmission can be seen both in quasilinear and nonlinear cases.

\section{Summary and conclusion}
In this paper we studied the nature of a globally propagated EUV wave associated with an X5.4 class flare occurred at AR11429 on March 7, 2012.
The X5.4 class flare started at 00:02UT and peaked at 00:24UT. A dome-like structure was observed in the images taken by {\it SDO}/AIA at 00:18UT and continued to expand. Then, at 00:23UT, another disturbance front appeared in the lower corona, and propagated toward the north with the velocity of around 670~km~s~$^{-1}$.
The foot print reached the polar prominence at 00:33UT.
On the other hand, in the images taken by {\it STEREO}/COR1, a disturbance appeared at 00:25UT around the expanding CME bubble, and passed the north pole between 00:31UT and 00:36UT. The appearance of the disturbance agreed well with the start of detection of Type II radio burst by HIRAS. The time variation of the location of the EUV disturbance which propagated to the north also agreed well with the location of the disturbance seen in the {\it STEREO}/COR1 images. 
From these observational properties of the EUV waves, we consider the foot print as a footprints of shock front launched from expanding CME bubble.

When the EUV wave hit the polar prominence, it is strongly brightened and the bright part propagated in the direction of EUV wave propagation. At the same time, the prominence was accelerated toward the direction of propagation of the wave. We consider the prominence activation as a result of shock injection. The strong brightening of the prominence could be explained as a result of the interaction of the shock with the prominence and resulting enhancement of the compression ratio of the prominence material. 

Using the observed velocity of activated prominence, the fast mode mach number $M_f$ of the coronal shock wave was estimated to be between 1.20 and 1.42 on the basis of linear theory. Also, linear theory predict the enhancement of the shock compression in the prominence material, which could explain the strong brightening of activated prominence in AIA 304~{\AA} images. The estimated compression ratio of the shock in the prominence was between 1.74 and 2.50. We checked the validity of the method by one dimensional numerical model calculation, and found that the linear theory is applicable when the shock is not strong.

\acknowledgements
The SDO/AIA data are courtesy of NASA/SDO and AIA science team. The authors would also like to thank SMART team at Hida observatory of Kyoto university who continuously provide high quality H$\alpha$ full disk image of the sun. The simulation code used in this work is created with the help of HPCI Strategic Program. This work is financially supported by JSPS KAKENHI Grant Numbers 24740331, 23340045 and 25287039. The autors are grateful to Dr. Andrew Hillier (Kyoto University) for helpful comments.

\clearpage

\begin{figure}
\epsscale{.90}
\plotone{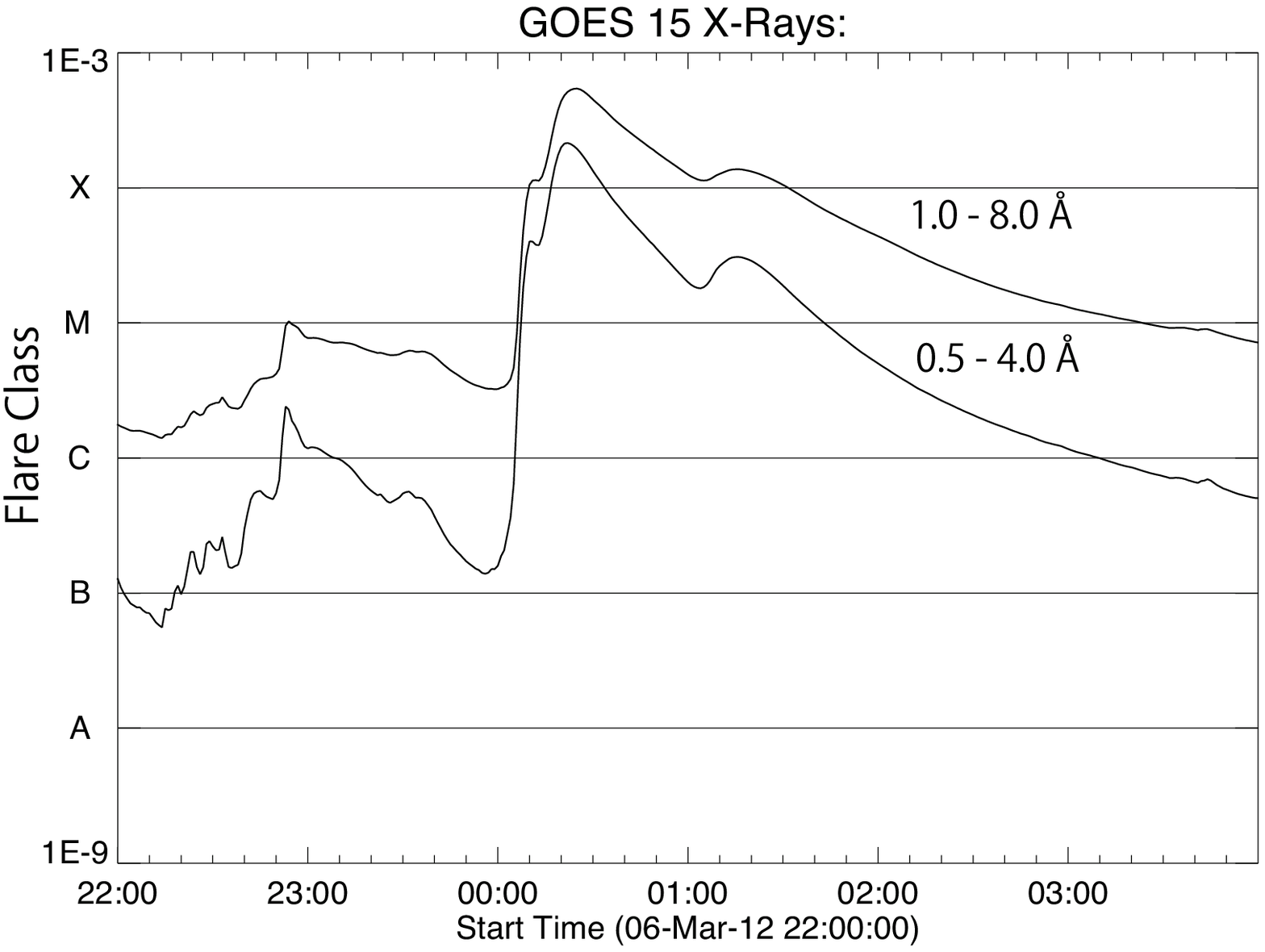}
\caption{GOES Soft X ray light curve.
X5.4 class flare occurred at 00:04UT and peaked at 00:24UT. One hour later X1.3 flare occurred at different part of the same active region.
\label{flare}}
\end{figure}

\begin{figure}
\epsscale{.90}
\plotone{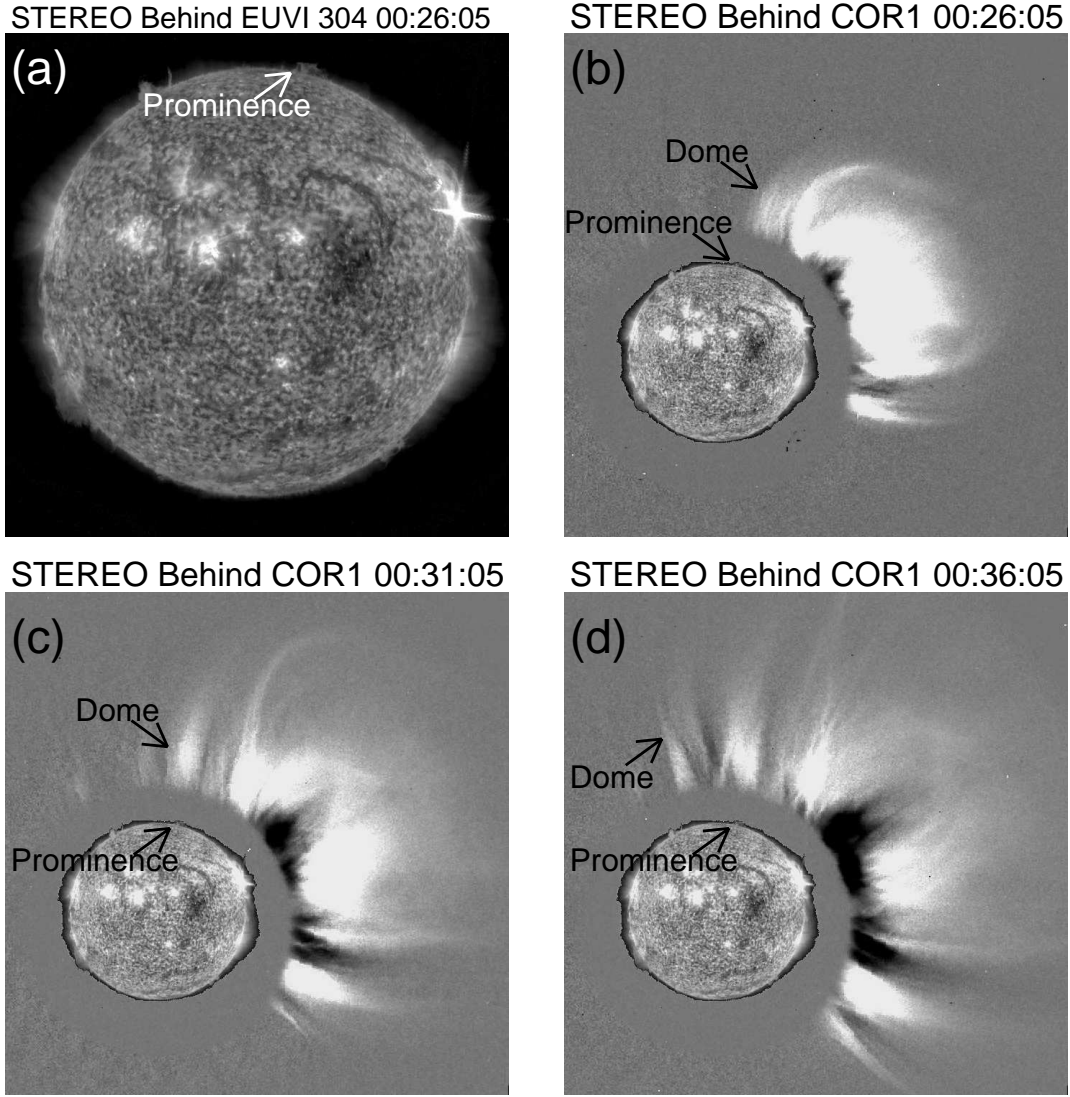}
\caption{Side view of the sun where we can see polar prominence and expanding dome-like structure.
Images are taken by STEREO Behind telescope. (a): EUVI 304~{\AA} image. We can see the polar prominence clearly. (b)-(d): Difference image of COR1 image with EUVI 304~{\AA} image embedded. Embedded 304~{\AA} image is the same as (a). The disturbance ahead of expanding CME appeared at 00:26UT and propagated to the north and passed the sky over the polar prominence.
\label{cme}}
\end{figure}

\begin{figure}
\epsscale{.80}
\plotone{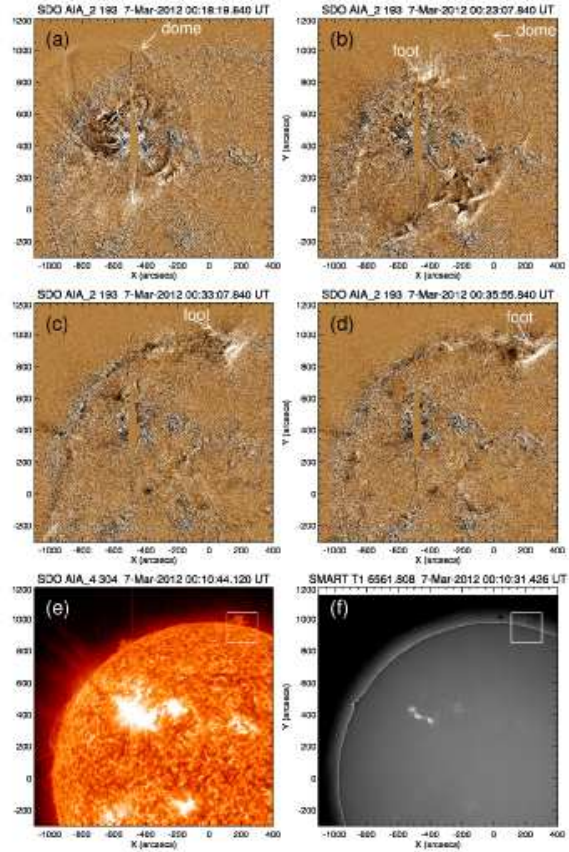}
\caption{(a)-(d): SDO/AIA 193~{\AA} difference image from 00:18UT to 00:36UT. (a): Dome-like structure appeared at 00:18UT. (b):foot print appeared at 00:23UT. (c): The foot print hit the polar prominence at 00:33UT. (d): foot print passed the polar prominence at 00:36UT. (e)AIA~304 image at 00:10UT.  We can identify the polar prominence clearly. (f): H${\alpha}$ image taken by SMART in Hida observatory. We can see the H${\alpha}$ ribbons ob the flare and polar prominence. The white rectangles in (e) and (f) correspond to the FOV of Figure 5.
\label{euv}}
\end{figure}

\begin{figure}
\epsscale{.90}
\plotone{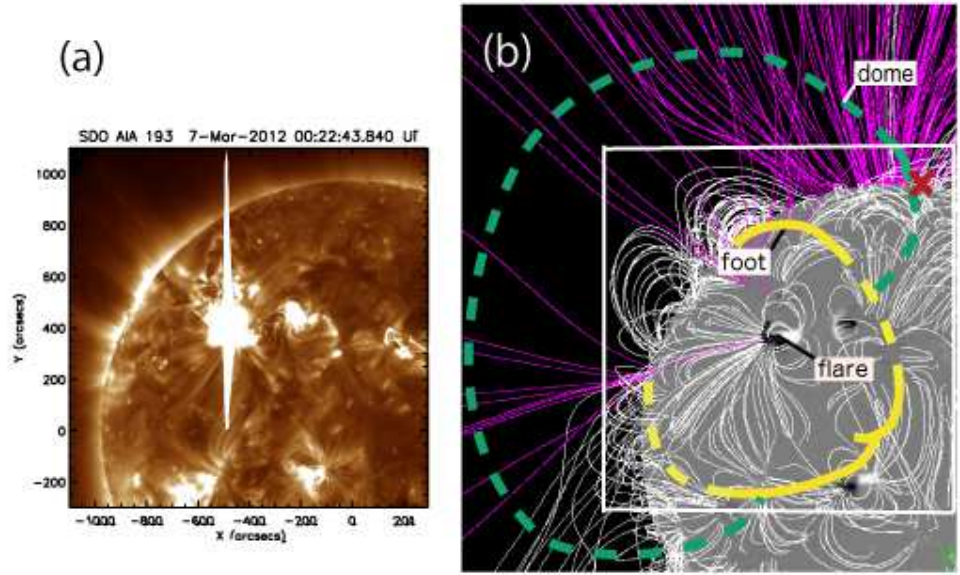}
\caption{(a): SDO/AIA 193~{AA} image taken at 00:22UT. (b): Schematic figure of EUV disturbance seen in SDO/AIA 193~{\AA} image (a).
The line of sight magnetic field obtained with SDO/HMI before the flare and the extrapolated coronal magnetic field (potential magnetic field) is shown. The red cross expresses the location of the prominence.
\label{schm}}
\end{figure}

\begin{figure}
\epsscale{.90}
\plotone{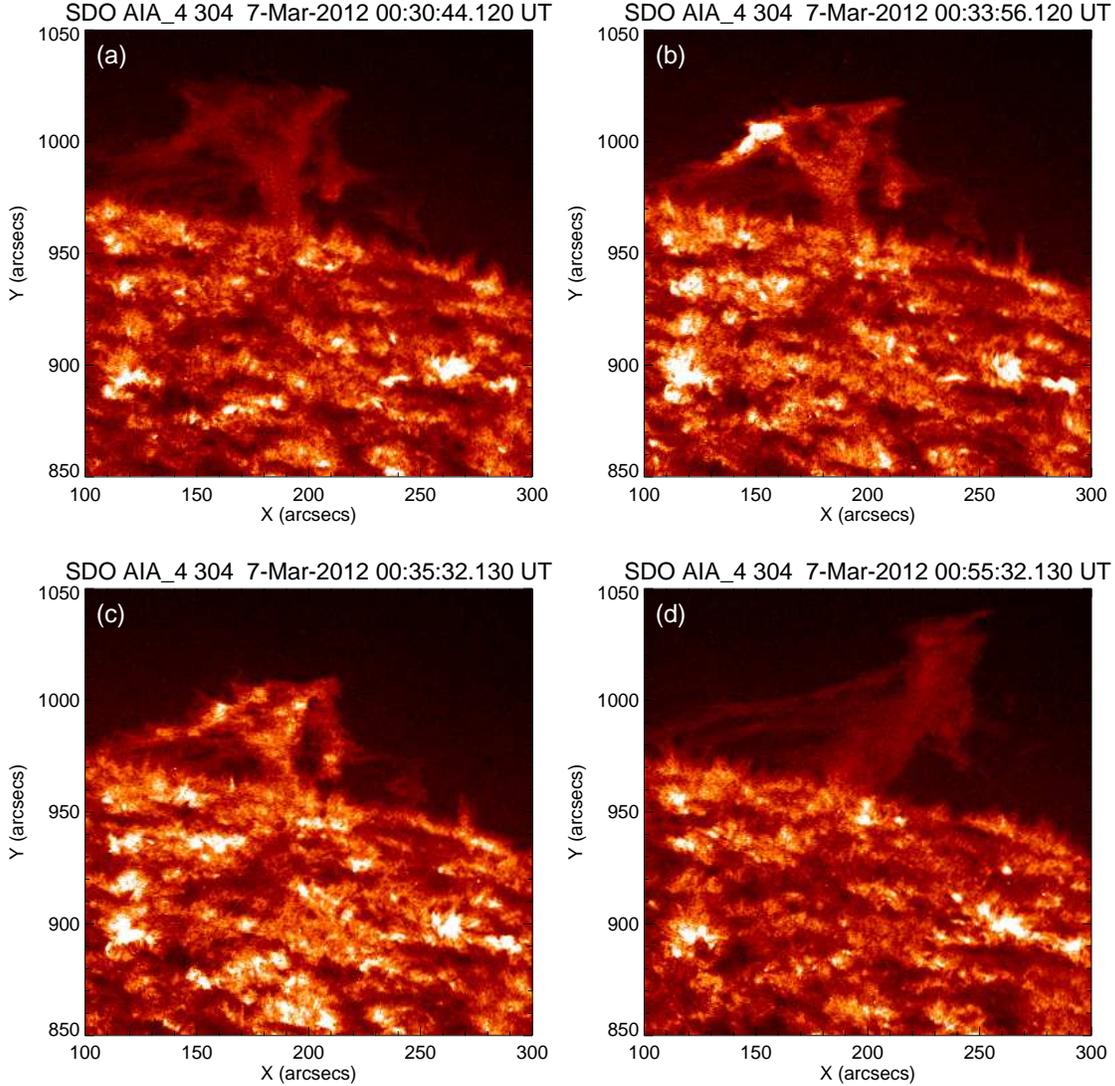}
\caption{AIA 304~{\AA} image of prominence activation.
(a): Before the arrival of the foot print at 00:31UT. (b): Just when the foot print reached the polar prominence at 00:34UT. The edge of the prominence strongly brightened. (c): While the acceleration of the prominence by the foot print at 00:36UT. A bright front propagates in the prominence toward the right. (d):When the prominence reached the maximum displacement at 00:55UT after the prominence activation had completed. The prominence kept oscillating after the activation had been completed.
\label{activation}}
\end{figure}

\begin{figure}
\epsscale{.90}
\plotone{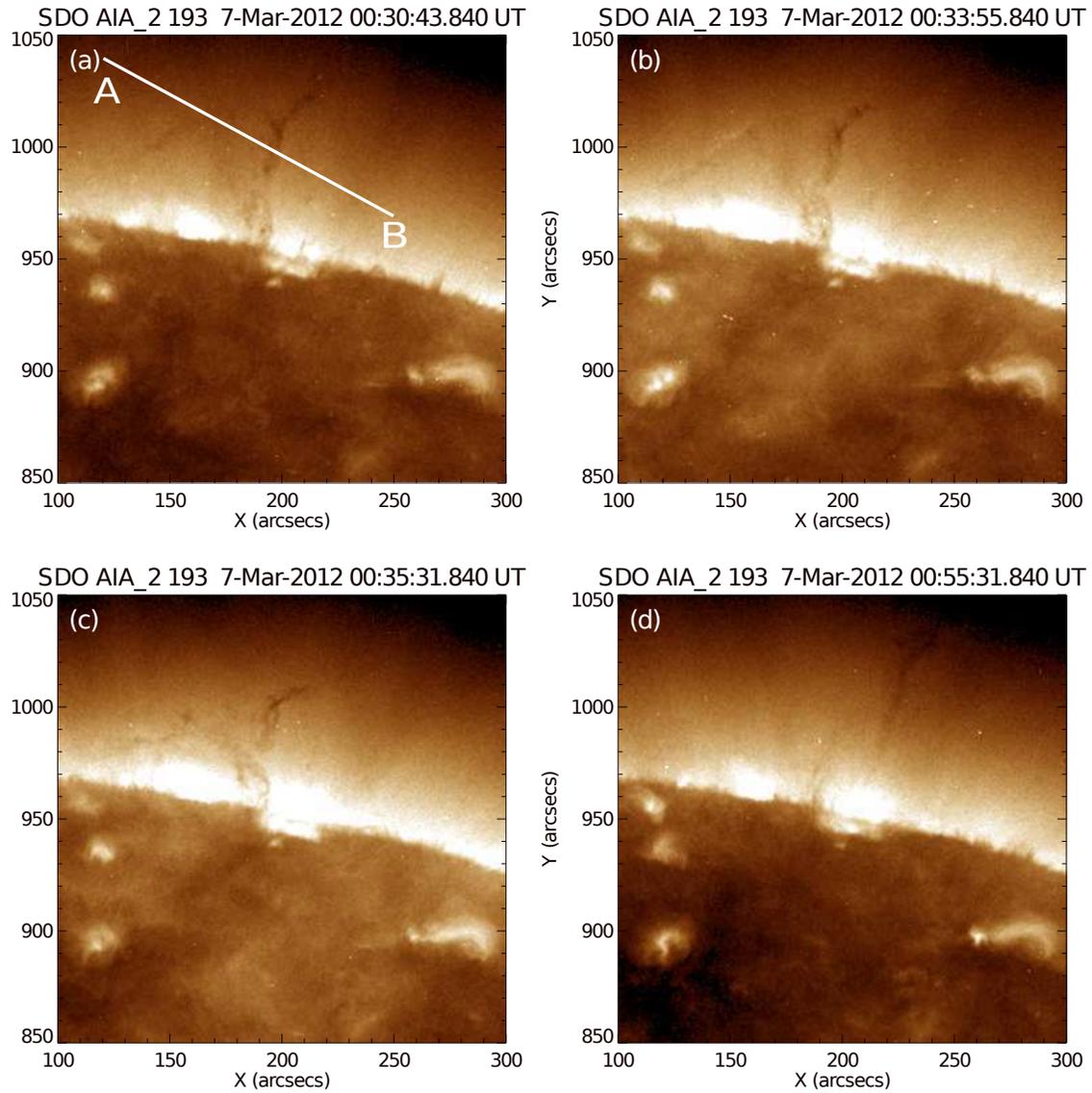}
\caption{AIA 193~{\AA} image of prominence activation. The FOV and timing of images (a)-(d) correspond to those of Figure 5 (a)-(d). The core of the polar prominence appears dark in 193~{\AA} pass band.
(a): Before the arrival of the foot print at 00:31UT. (b): Just when the foot print reached the polar prominence at 00:34UT. (c): While the acceleration of the prominence by the foot print at 00:36UT.  (d): When the prominence reached the maximum displacement at 00:55UT after the prominence activation had completed. The prominence kept oscillating after the activation had been completed.
\label{activation}}
\end{figure}

\begin{figure}
\epsscale{.90}
\plotone{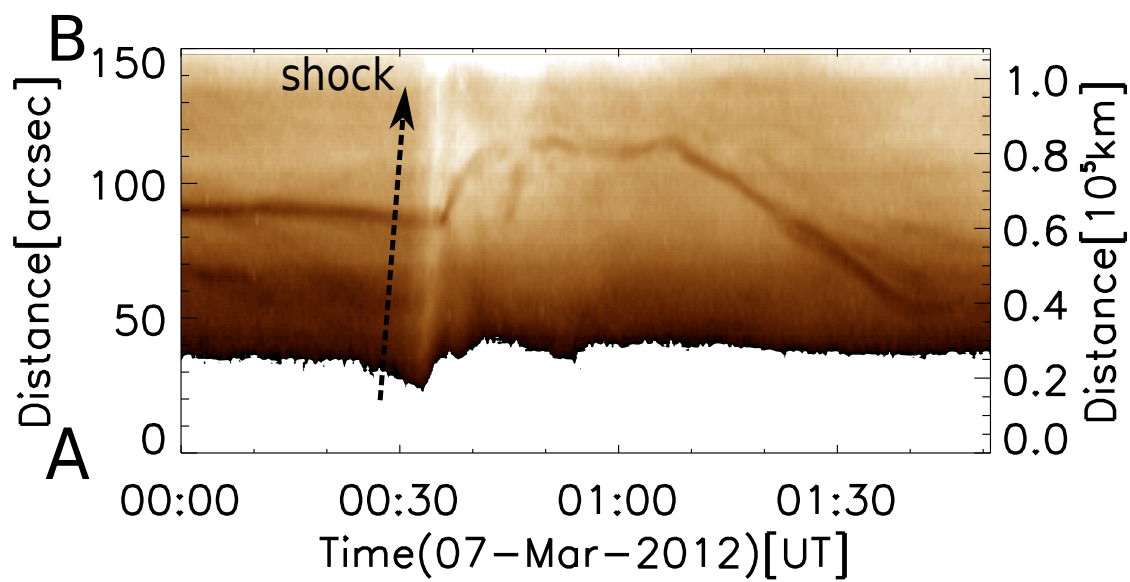}
\caption{AIA 193~{\AA} time-distance plot of prominence activation along the line A-B in Figure 7. The pixels whose pixel count is lower than the threshold value which is set to emphasize both the signatures of wave front and dark filament is out of color table and appears white in the figure. The bright feature approaching the dark filament is the coronal shock front. We can see that the propagation speed of the shock front does not change much in the corona around the prominence. Just when the shock front arrives at the prominence, the prominence is suddenly accelerated. After the activation, the prominence reach its maximum displacement and continues oscillation.
\label{wave_activation}}
\end{figure}

\clearpage

\begin{table}
\caption{Estimated shock properties for $\gamma = 5/3$ \newline ~~~~~~~~~~~~on the basis of the linear theory \label{tbl-1}}
\begin{tabular}{crrrrrrrrrrr}
\tableline
\tableline
\multicolumn{1}{c}{$f_V$\tablenotemark{a}} & \multicolumn{1}{c}{$\beta_c$\tablenotemark{b}} &
 \multicolumn{1}{c}{$r_c$\tablenotemark{c}} & \multicolumn{1}{c}{$r_p\tablenotemark{d}$} & \multicolumn{1}{c}{$M_{f,c}$\tablenotemark{e}}\\
\tableline
0.001. &$~~~$  0.05 &$~~~$  1.15 &$~~~$  1.18 &$~~~$  1.11$~~~~$ \\
0.001 &$~~~$  0.20 &$~~~$  1.15 &$~~~$  1.18 &$~~~$  1.11$~~~~$ \\
0.01 &$~~~$  0.05 &$~~~$  1.18 &$~~~$  1.38 &$~~~$  1.14$~~~~$ \\
0.01 &$~~~$  0.20 &$~~~$  1.18 &$~~~$  1.38 &$~~~$  1.13$~~~~$ \\
0.1 &$~~~$  0.05 &$~~~$  1.37 &$~~~$  2.11 &$~~~$  1.29$~~~~$ \\
0.1 &$~~~$  0.20 &$~~~$  1.37 &$~~~$  2.11 &$~~~$  1.29$~~~~$ \\

\tableline
\end{tabular}
\tablenotetext{a}{The filling factor of the prominence}
\tablenotetext{b}{Plasma beta in the corona}
\tablenotetext{c}{Compression ratio of the shock wave in the corona}
\tablenotetext{d}{Compression ratio of the shock wave in the prominence}
\tablenotetext{e}{Fast mode Mach number of the shock wave in the corona}
\end{table}

\clearpage
\begin{figure}
\includegraphics{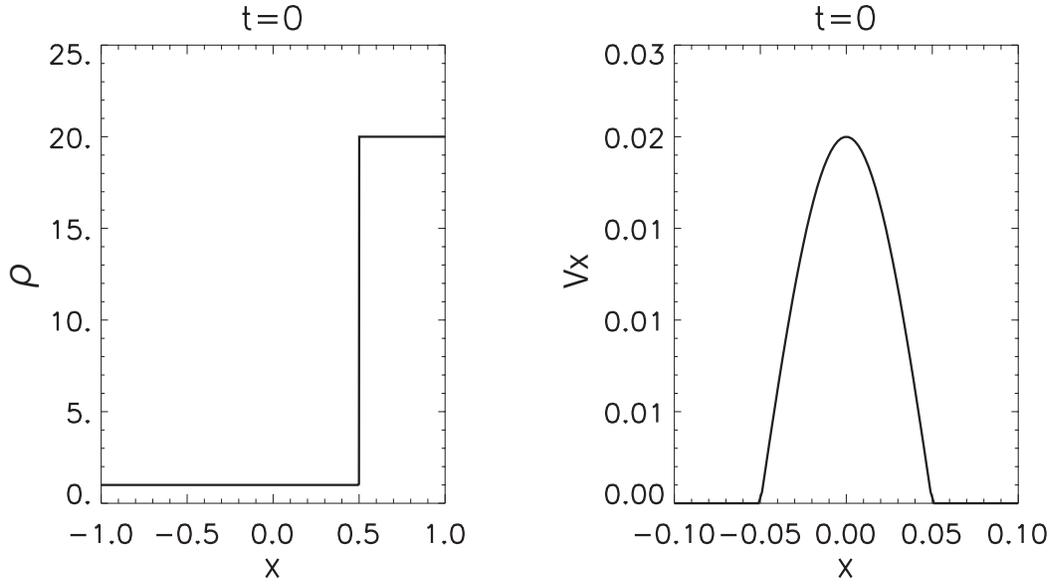}
\caption{Initial profile of the density(left) and velocity(right) for a quasilinear case with $\rho_{p}=20$.
\label{simulation}}
\end{figure}

\begin{figure}
\includegraphics{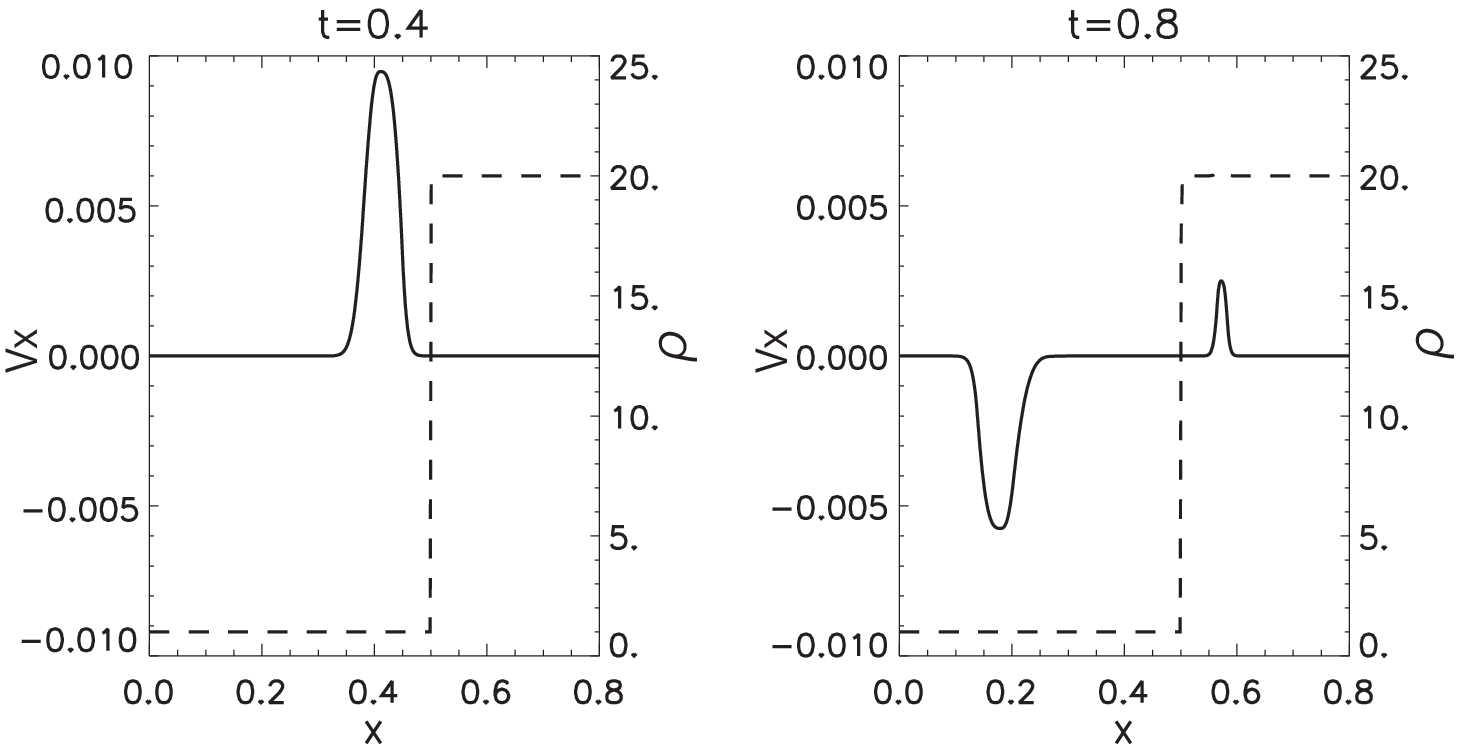}
\caption{Vx just before and after the wave transmission in the quasilinear case.
Velocity amplitudes of injected ($V_i$) and transmitted ($V_t$) waves are indicated.
\label{simulation}}
\end{figure}

\begin{figure}
\includegraphics{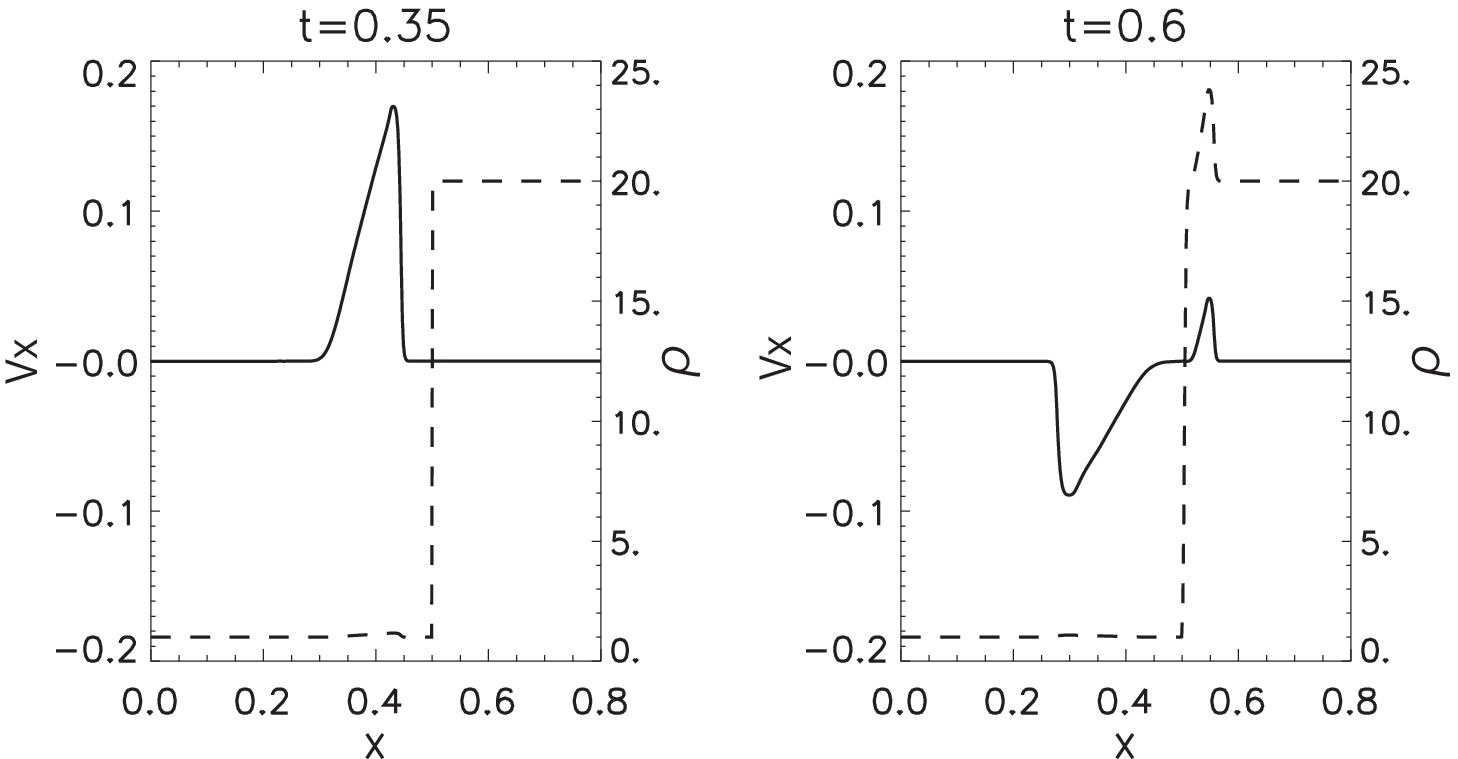}
\caption{Vx just before and after the wave transmission in the nonlinear case.
Velocity amplitudes of injected ($V_i$) and transmitted ($V_t$) waves are indicated.
\label{simulation}}
\end{figure}

\begin{figure}
\includegraphics{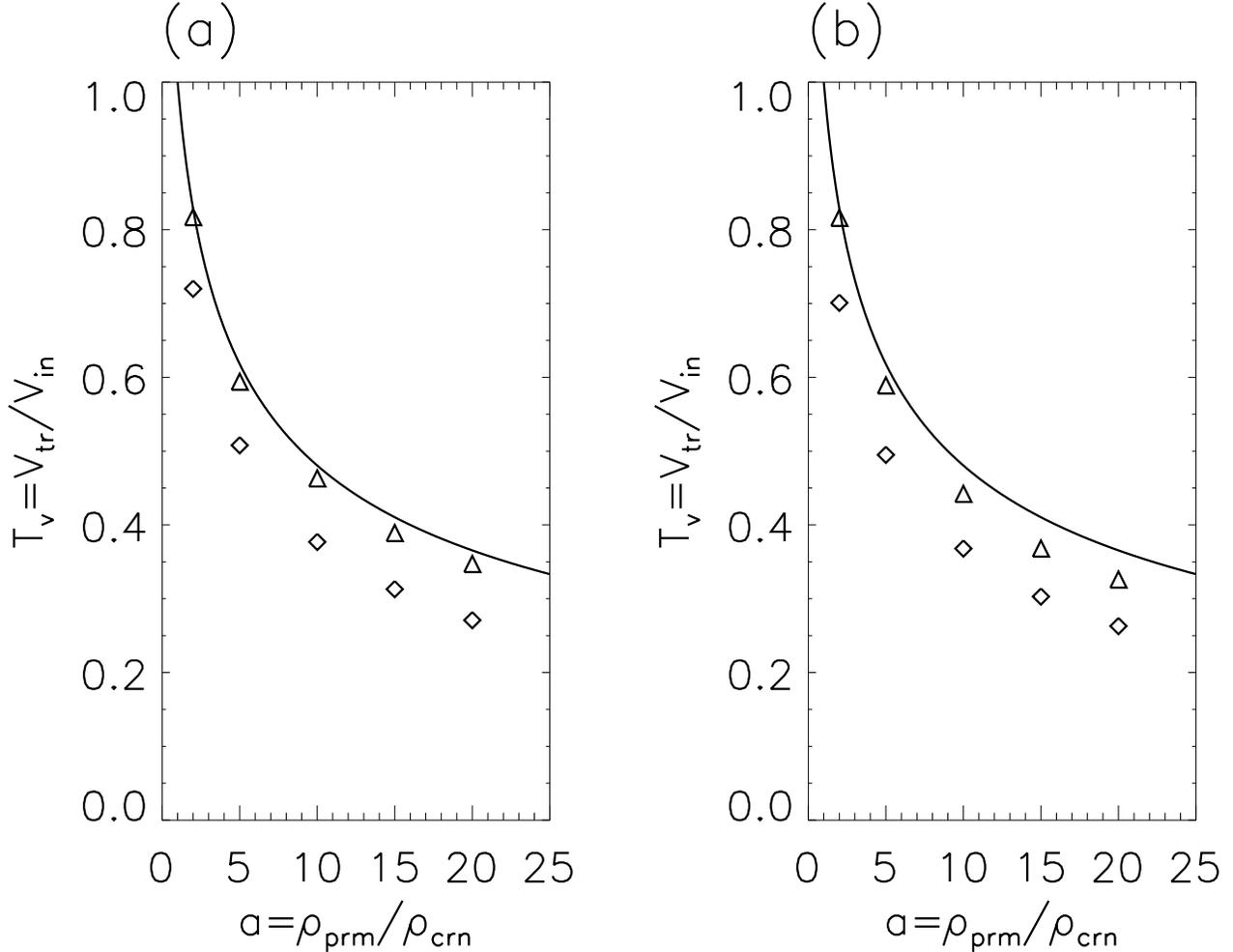}
\caption{The result of one dimensional simulation of shock transmission.
The velocity transmittance $T_v$ is shown as a function of density gap ${\rm a}$. Solid lines denote linear analytic solution, triangles and diamonds show the numerical results of quasilinear and nonlinear (shock) transmission cases, respectively. (a): Plasma beta $\beta=0.05$ case. Fast mode Mach number of the injected shock wave $M_f=1.16$. (b): Plasma beta $\beta=0.20$ case. Fast mode Mach number of the injected shock wave $M_f=1.17$. 
\label{simulation}}
\end{figure}

\begin{figure}
\includegraphics{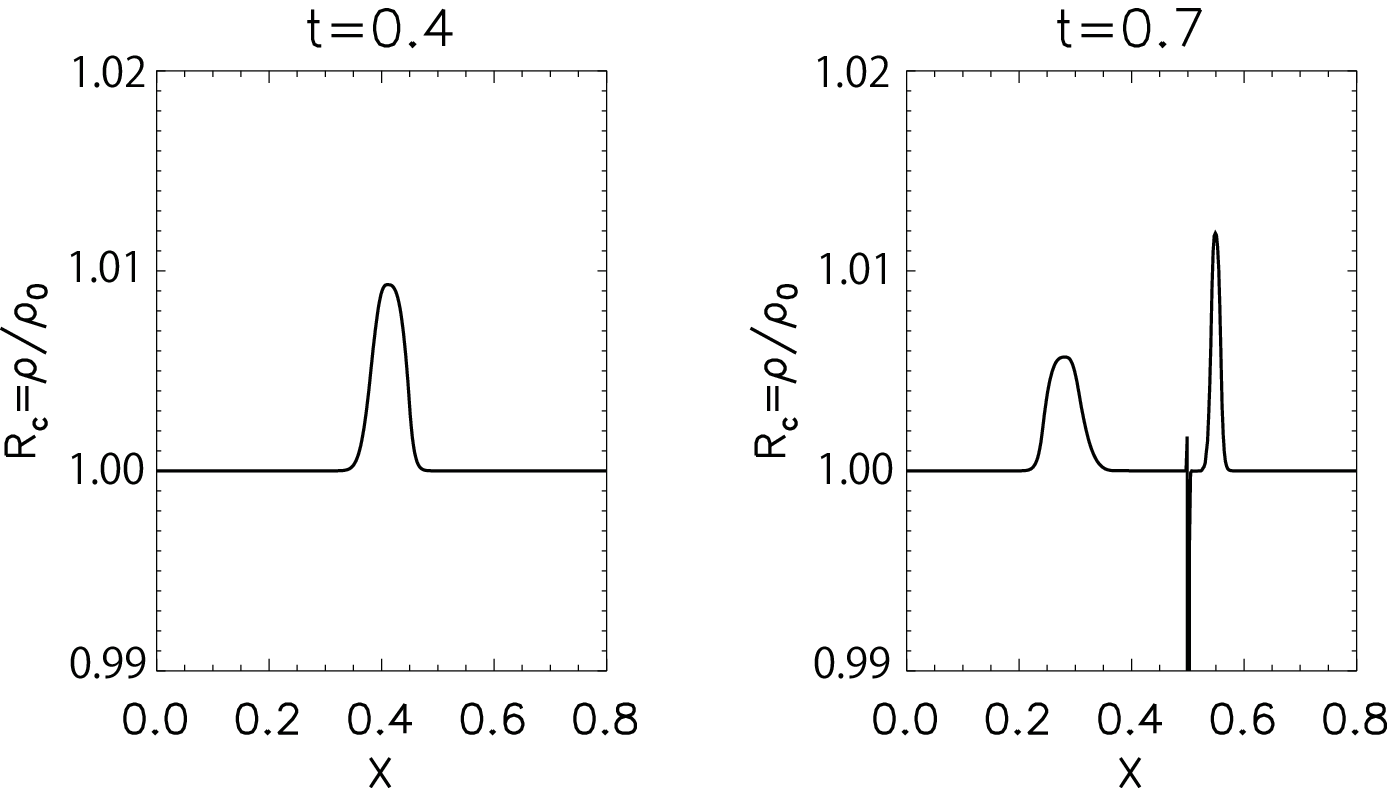}
\caption{$R_c$ just before and after the wave transmission in the quasilinear case.
Enhancement of $R_c$ after the transmission can be seen.
\label{simulation}}
\end{figure}

\begin{figure}
\includegraphics{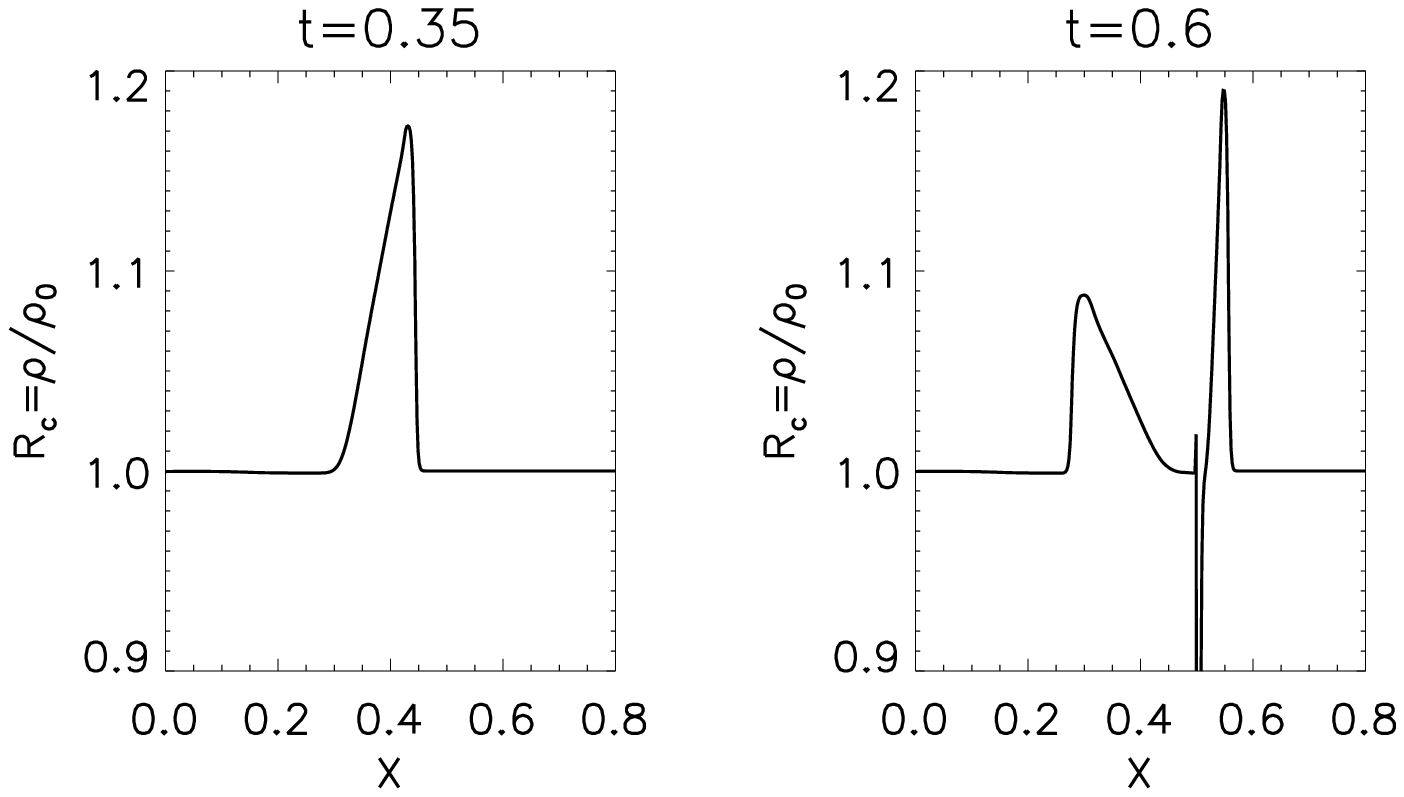}
\caption{$R_c$ just before and after the wave transmission in the nonlinear case.
Enhancement of $R_c$ after the transmission can be seen.
\label{simulation}}
\end{figure}

\begin{figure}
\includegraphics{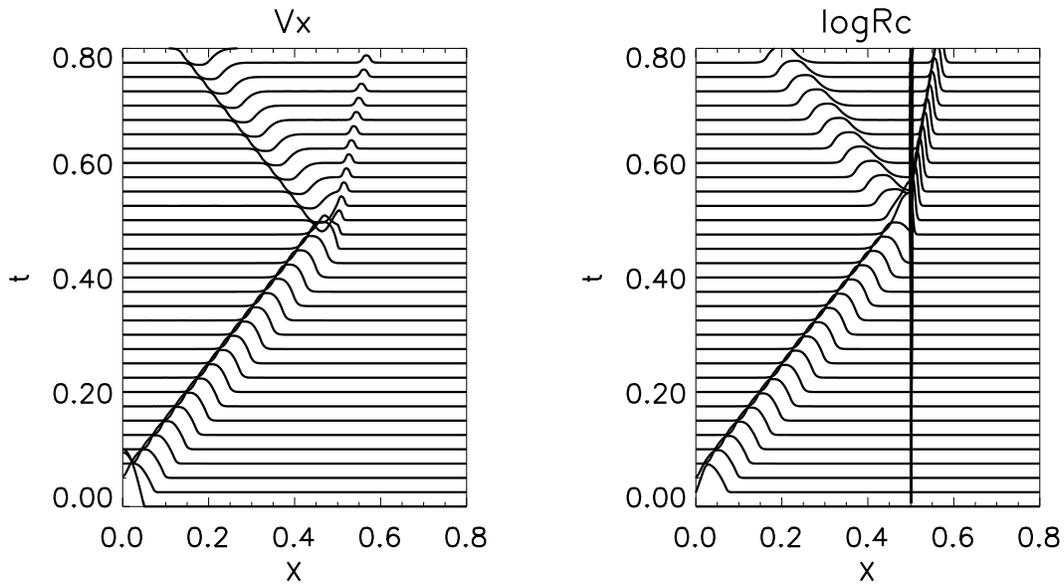}
\caption{Time evolution of $V_x$ and $log{R_c}$ in quasilinear case is shown.
\label{simulation}}
\end{figure}

\begin{figure}
\includegraphics{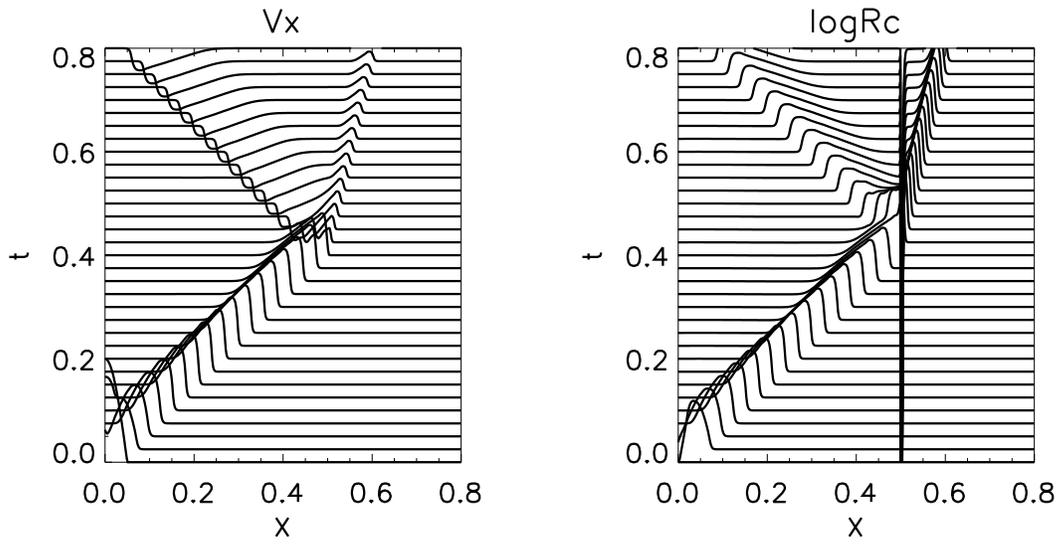}
\caption{Time evolution of $V_x$ and $\log{R_c}$ in nonlinear case is shown.
\label{simulation}}
\end{figure}

\clearpage

\end{document}